\documentclass{IEEE_lsens}
\usepackage{graphicx}
\usepackage{multirow}

\usepackage{epsfig} 
\usepackage{mathptmx} 
\usepackage{times} 
\usepackage{amsmath} 
\usepackage{amssymb}  
\usepackage{flushend}
\usepackage{balance}
\usepackage{units}
\usepackage{cite}
\usepackage{upgreek}
\usepackage{subfloat}
\usepackage[caption=false]{subfig}
\usepackage{xcolor}
\usepackage{tabularx}
\newcolumntype{C}[1]{>{\centering\arraybackslash}p{#1}}  
\newcolumntype{R}[1]{>{\raggedleft\arraybackslash}p{#1}}  
\newcolumntype{L}[1]{>{\raggedright\arraybackslash}p{#1}}  
\renewcommand\tabcolsep{3pt}
\begin{document}
\markboth{Vol., No., ~2018}{0000000}
\IEEELSENSarticlesubject{Sensor Applications}
\title{Power Distribution System Synchrophasors with Non-Gaussian Errors: Real-World Measurements and Analysis}
\author{\IEEEauthorblockN{Can Huang\IEEEauthorrefmark{}, Charanraj A. Thimmisetty\IEEEauthorrefmark{}, Xiao Chen\IEEEauthorrefmark{}, Mert Korkali\IEEEauthorrefmark{}, Vaibhav Donde\IEEEauthorrefmark{}, Emma Stewart\IEEEauthorrefmark{}, Philip Top\IEEEauthorrefmark{}, Charles Tong\IEEEauthorrefmark{}, Liang Min\IEEEauthorrefmark{}}
\IEEEauthorblockA{\IEEEauthorrefmark{}Lawrence Livermore National Laboratory, Livermore, CA 94550, USA}

\thanks{Corresponding author: X. Chen (e-mail: chen73@llnl.gov).}
\thanks{Associate Editor: }
\thanks{Digital Object Identifier: }
}

\IEEEtitleabstractindextext{%
\begin{abstract}

This letter studies the synchrophasor measurement error of electric power distribution systems with on-line and off-line measurements using graphical and numerical tests. It demonstrates that the  synchrophasor measurement error follows a non-Gaussian distribution instead of the traditionally-assumed Gaussian distribution. It suggests the need to use non-Gaussian or Gaussian mixture models to represent the  synchrophasor measurement error. These models are more realistic to accurately represent the error than the traditional Gaussian model. The measurements and underlying analysis will be helpful for the understanding of distribution system measurement characteristics, and also for the modeling and simulation of distribution system applications.

\end{abstract}

}
\maketitle



\section{Introduction}

\IEEEPARstart An electric power grid is an interconnected network for delivering electricity from generators to loads via transmission and distribution systems. It is also a network overlaid with sensing and measurement, communication, and monitoring and control components that maintain grid reliability, security, and efficiency. Today's power grid has been evolving into the `smart grid' to provide more reliable, more efficient, and more sustainable electricity to customers~\cite{IEEE, S1, S2}. To achieve these attributes, a variety of smart grid technologies are needed.

In sensor and measurement fields, the `synchrophasor' is one of the most important smart grid technologies~\cite{IEEE, S1,S2, S3, D1, D2}. A synchrophasor system consists primarily of phasor measurement units (PMUs), phasor data concentrators (PDCs), and communication networks, as shown in Fig. 1. It typically uses PMUs to produce synchrophasor measurements from current and voltage signals (e.g., the ones from current and voltage transducers) and a standard time signal (e.g., the one from a global positioning system (GPS)). It then utilizes PDCs to transfer synchrophasor data from PMUs/PDCs to a control center and/or various applications~\cite{S1,S2, S3, IEEE}. In the past decade, an increasing number of synchrophasor systems have been installed around the world and a series of synchrophasor applications have been implemented in grids. The North American SynchroPhasor Initiative (NASPI) reports that there are about 2,000 commercial PMUs installed across North America, and more than 20 kinds of PMU-based applications under research and development~\cite{D1, D2}.
\begin{figure}[h]
\centering
{\includegraphics[width=1.0\linewidth]{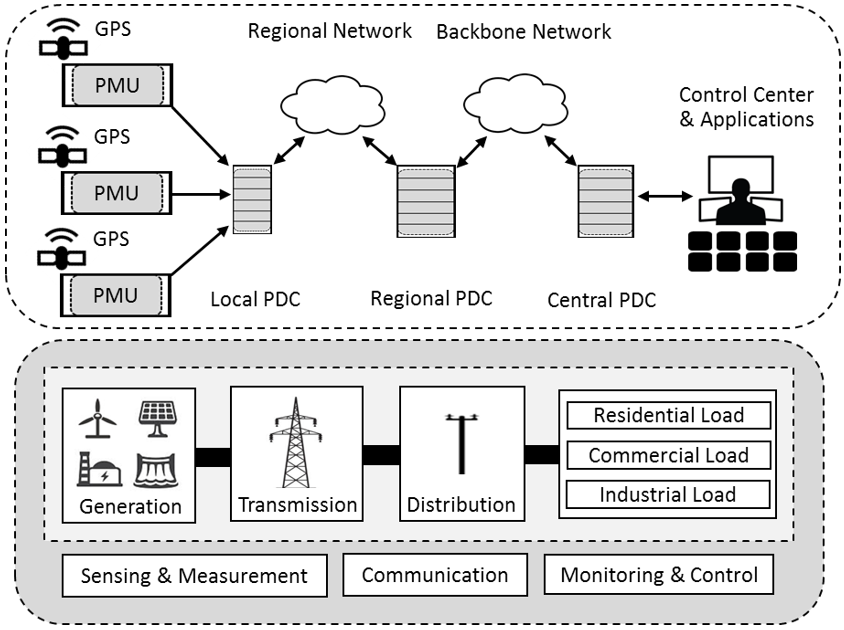}}
\caption{A synchrophaor system over an electric power grid.}\label{fig:pdf}
\end{figure}

The synchrophasor is expected to perform high-precision, low-latency, and time-synchronized measurement and provide significant insight into grid planning and operation. In practice, the synchrophasor inevitably involves measurement errors, which may affect or even disable certain synchrophasor applications~\cite{D1, D2, S4, Zhao1,Zhao2,Zhao3}. It is a challenging yet critical task to analyze and model the synchrophasor measurement error. Traditionally, the synchrophasor is designed for transmission systems and the synchrophasor measurement error is assumed as a Gaussian noise in most synchrophasor applications. Several studies point out that this assumption is violated in reality and the results are misleading or even damaging to certain applications (e.g., PMU-based state estimation)~\cite{Zhao1,Zhao2,Zhao3}. For example, Wang \textit{et al}. analyze real PMU measurements and reveal that the PMU measurement errors do not follow a Gaussian distribution~\cite{Zhao1}; and Mili \textit{et al}. assess the sensitivity of different state estimators to Gaussian/non-Gaussian noises and develop a robust state estimation method to cope with non-Gaussian measurement errors~\cite{Zhao2}. 

Now the synchrophasor is being extensively deployed in distribution systems, such as micro-PMUs ($\mu$PMUs) and FNET/GridEye~\cite{NASPI, PSL, von2017precision, FNET}. Compared with transmission systems, distribution systems are invested with fewer sensors and measurements, and distribution networks are often plagued by large measurement uncertainties due to their highly distributed and diverse infrastructure. At present we pay increasingly more attention to distribution systems, especially with the rapid development of distributed energy resources (DER) and distribution management systems (DMS). However, we have very limited knowledge about the nature of distribution system measurement errors. Accordingly, NASPI established Distribution Task Team (NASPI DisTT) to promote the distribution-level synchrophasor development. In NASPI DisTT 2017 Winter Report, one of the most urgent tasks is to investigate the nature of the synchrophasor measurement error in real distribution systems~\cite{NASPI}. 

This letter investigates the distribution system synchrophasor measurement error with on-line and off-line measurements, and identifies the distribution of the measurement error using both graphical and numerical methods. To the best knowledge of the authors, this is the first paper to perform this kind of studies with real-world measurements and analysis. The results will be useful for the modeling and simulation of distribution systems, and also for the design and development of advanced DER and DMS applications. 

\section{Methodology}
In this letter, the measurement error is defined as the difference between the measured and true values of a selected quantity. It mainly consists of two components: a systematic error and a random error, which are often represented by a consistent bias and a random noise, respectively. In theory, the random error plays a decisive role in the distribution of measurement errors, which is typically introduced by unknown and unpredictable changes occurring in measurement devices (e.g., the electronic noise and circuit aging) and/or in the environment (e.g., the wind, temperature, and communication). To investigate the distribution of  the distribution-level synchrophasor measurement error thoroughly, tests are performed in various measurement devices including PMUs, $\mu$PMUs, and FNET/GridEye, and in different environments covering both primary and secondary distribution systems. In the following, we begin with the explanation of the measurement and analysis methods.

\paragraph*{\textbf{Measurement}} 
It is difficult to obtain directly the real-time measurement error between the measurement and the true values. Here, the distribution of the measurement error is identified indirectly through multiple  synchronized measurements (MSMs), whereby multiple identical and independent  synchrophasor measuring devices are deployed to simultaneously and independently meter the quantity $x$ with the measurement ($z_i = x + e_i$, $z_j = x + e_j$, $i$ and $j$ $\in N$) and the measurement error ($e_i$, $e_j$). Subsequently, the Gaussian/non-Gaussian distribution of the measurement error can be determined by constructing the distribution of the difference of the MSM errors or MSMs, i.e.  $\Delta e = e_i - e_j$, $\Delta z= z_i - z_j$, and $\Delta e =\Delta z$. The general principle is: if A and B are two independent random variables, and they are both normally distributed, then $\text{A} \pm \text{B}$ is normally distributed (\textbf{Proposition}); and if $\text{A} \pm \text{B}$ is not normally distributed, then either A or B (or both) is not normally distributed (\textbf{Contraposition}). In other words, if $\Delta e = e_i - e_j$ is non-Gaussian, then $e_i$ or/and $e_j$ is non-Gaussian. Since the measurements are taken from identical devices around very similar environment where the measurement errors are supposed to follow the same distribution, it can thus be argued that if $\Delta e $ is non-Gaussian, then both $e_i$ and $e_j$ are non-Gaussian. In addition to the on-line test described above, an off-line test is carried out on a distribution monitoring platform FNET/GridEye, in which a distribution signal is generated by a power system simulator and measured by a high-precision frequency disturbance recorder (FDR), and the resulting measurement error is calculated by a calibrator.

\paragraph*{\textbf{Analysis}}
\begin{figure}[h]
%
%
\centering
\subfloat{\includegraphics[width=0.32\linewidth]
{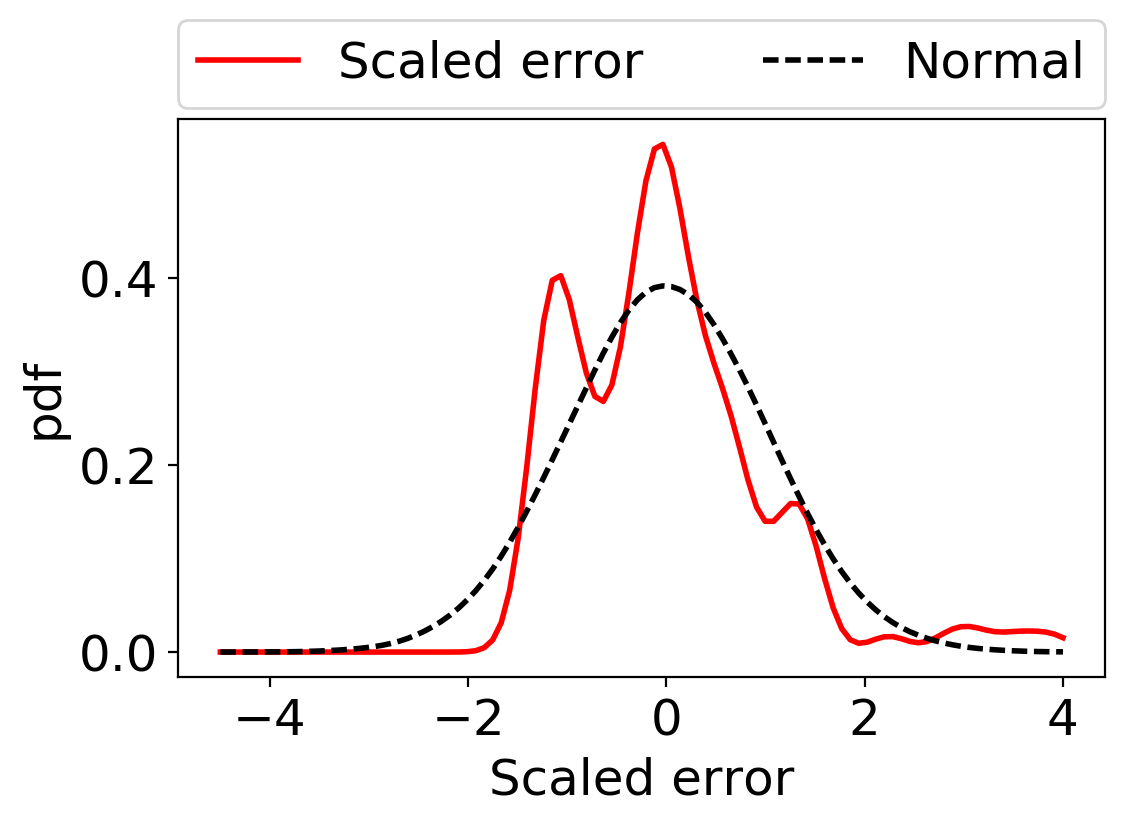}} \addtocounter{subfigure}{-1}
\subfloat[]{\includegraphics[width=0.32\linewidth]{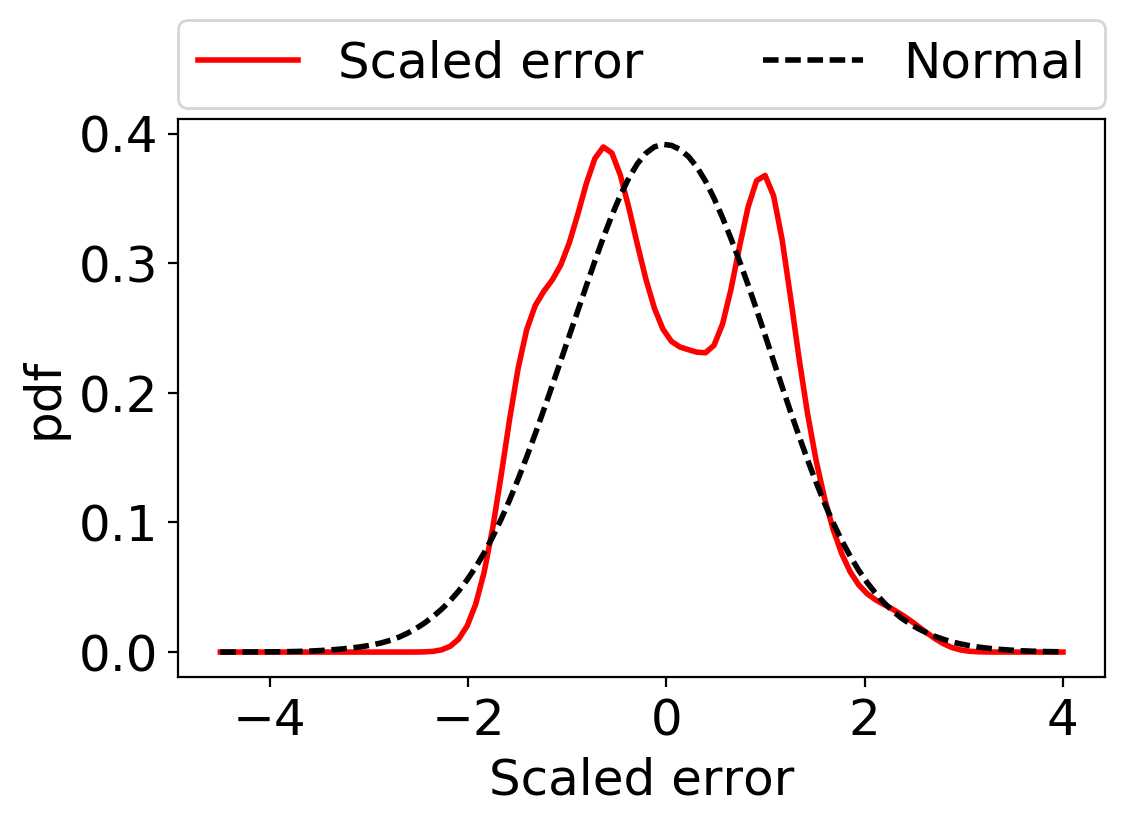}} \subfloat{\includegraphics[width=0.32\linewidth]{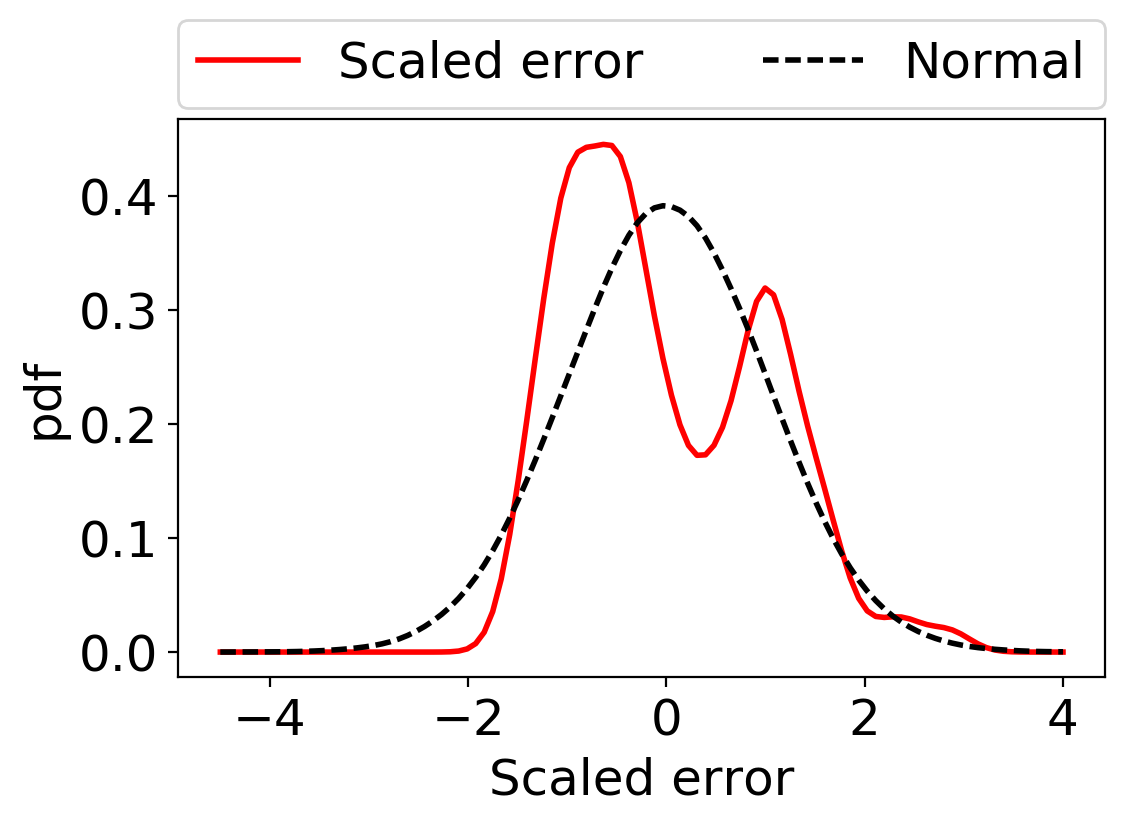}}\addtocounter{subfigure}{-1} \\
\vspace{-0.15in}
\subfloat{\includegraphics[width=0.32\linewidth]
{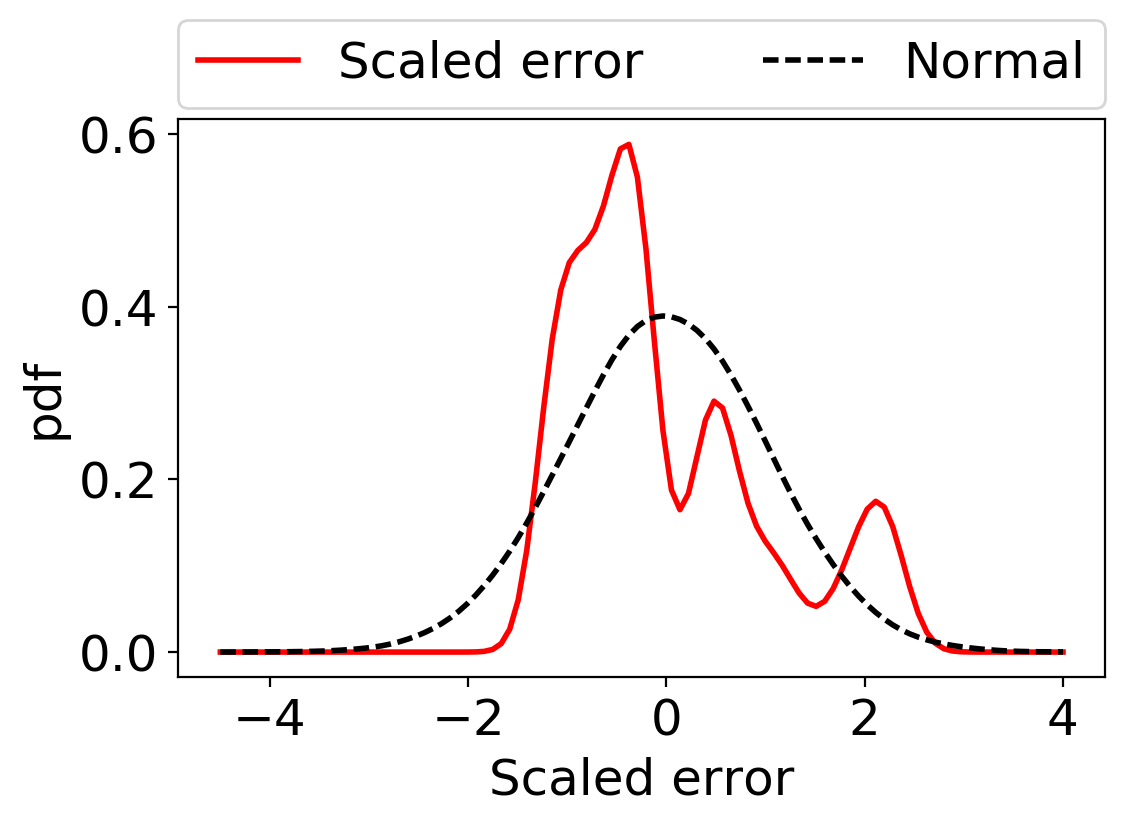}}\addtocounter{subfigure}{-1}
\subfloat[]{\includegraphics[width=0.32\linewidth]{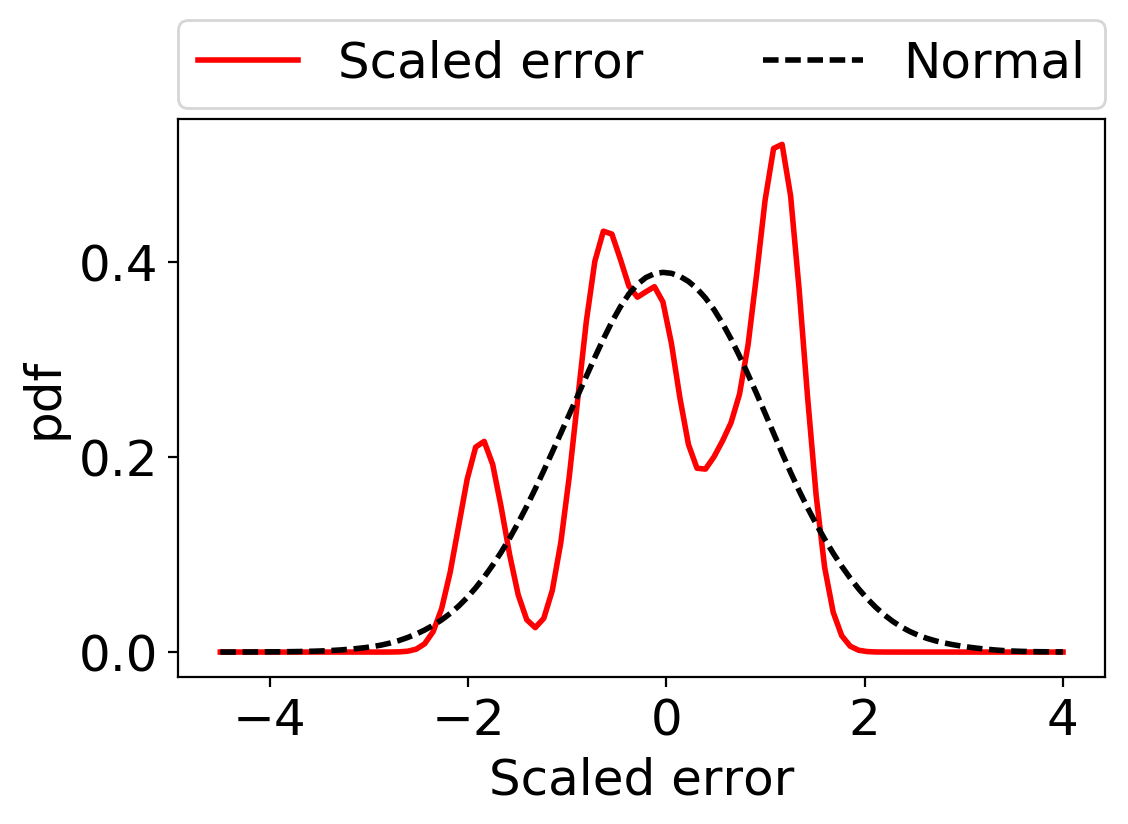}} 
\subfloat{\includegraphics[width=0.32\linewidth]{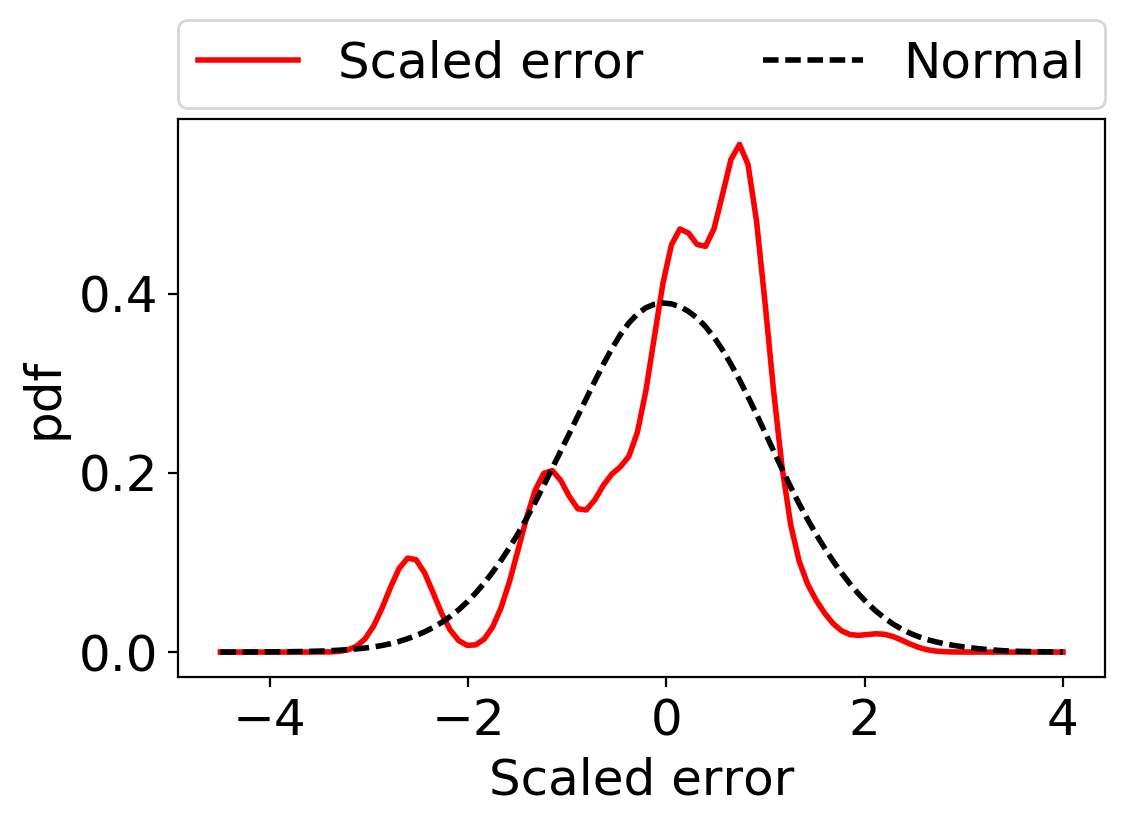}}\addtocounter{subfigure}{-1}\\
\vspace{-0.15in}
\subfloat[]{\includegraphics[width=0.5\linewidth]{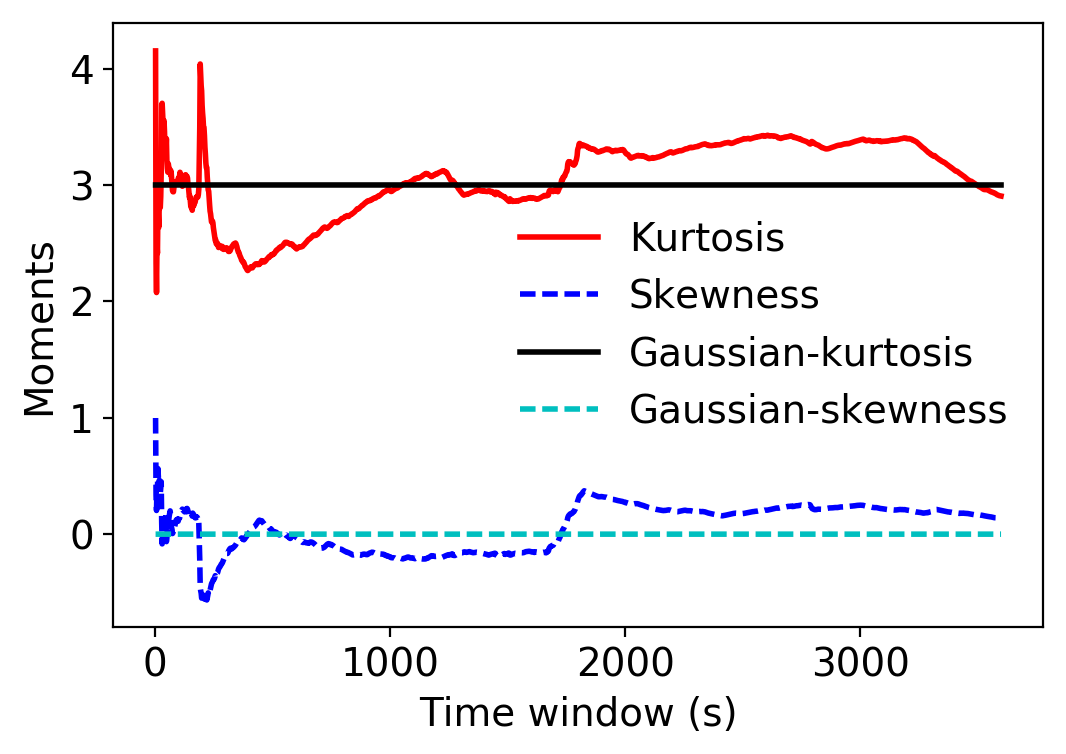}}
\subfloat[]{\includegraphics[width=0.5\linewidth]{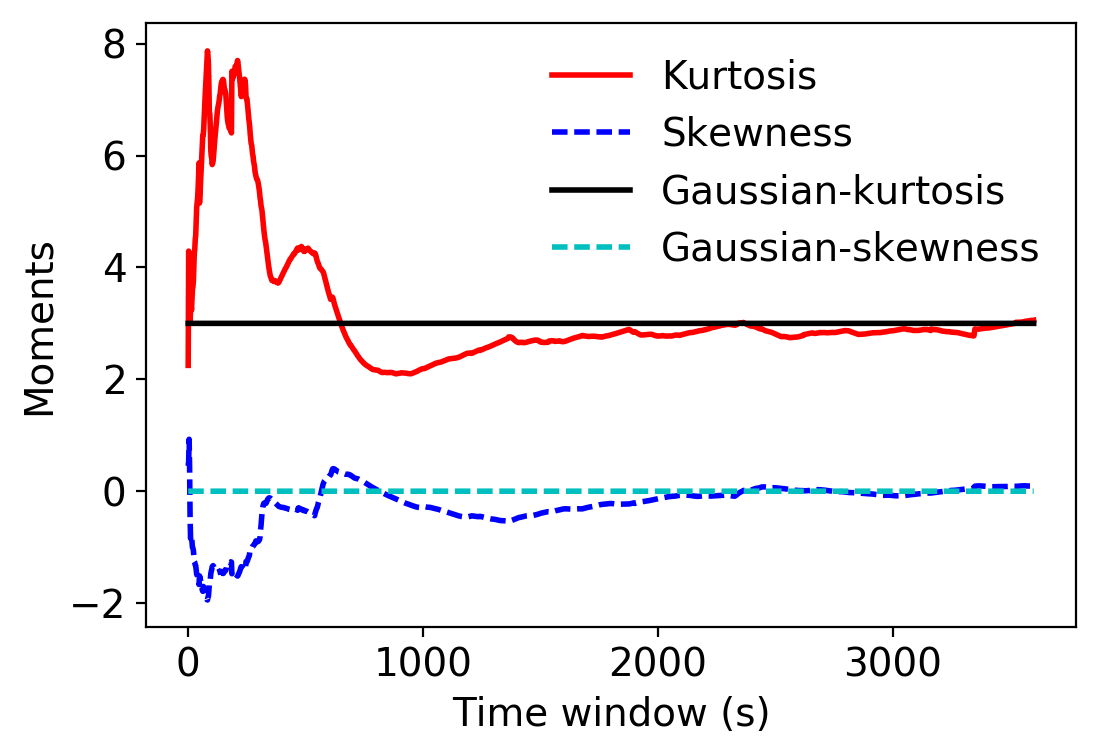}}
\caption{Test 1 results: (a) pdf of the voltage angle error differences $\Delta e_s$ with standard Gaussian for time windows of 1s, 5s, 10s; (b) pdf of the voltage magnitude error differences $\Delta e_s$ with standard Gaussian for time windows of 1s, 5s, and 10s; (c) varying  time windows versus skewness and kurtosis of voltage angle  $\Delta e$; and (d) varying update period versus skewness and kurtosis of voltage magnitude $\Delta e$.}\label{fig:pdf}
\end{figure}
\begin{table}[h]
\centering
\caption{Summary of the Gaussianity Tests on $\mu$PMU Data}\label{tab:Gper}
\begin{tabular}{|c|c|c|c|c|c|c|}
\hline
\multicolumn{1}{|c|}{\multirow{3}{*}{\begin{tabular}[c]{@{}c@{}}Measurement\end{tabular}}}&
\multicolumn{1}{c|}{\multirow{3}{*}{\begin{tabular}[c]{@{}c@{}}Time\\ window\\ (s)\end{tabular}}} &\multicolumn{1}{c|}{\multirow{3}{*}{\begin{tabular}[c]{@{}c@{}}Sample\\size \end{tabular}}} & \multicolumn{4}{c|}{\begin{tabular}[c]{@{}c@{}}\% of non-Gaussian distributions\end{tabular}}                                   \\ \cline{4-7} 
& & & \multicolumn{2}{c|}{Shapiro-Wilk}& \multicolumn{2}{c|}{Kolmogorov-Smirnov}\\ \cline{4-7} 
& & & \multicolumn{1}{c|}{$\alpha$=5\%} & \multicolumn{1}{c|}{$\alpha$=10\%} & \multicolumn{1}{c|}{$\alpha$=5\%} & \multicolumn{1}{c|}{$\alpha$=10\%} \\ 
\hline
\multicolumn{1}{|c|}{\multirow{4}{*}{\begin{tabular}[c]{@{}c@{}}Voltage\\angle\end{tabular}}}
&1 &120& 89.5 & 93.5& 40& 49.5 \\
&5 &600& 99& 99& 75&  81.5\\
&10 &1200& 100& 100& 76.5 &  84 \\
&30 &3600& 99.5& 100& 81&85\\
\hline
\multicolumn{1}{|c|}{\multirow{4}{*}{\begin{tabular}[c]{@{}c@{}}Voltage\\magnitude\end{tabular}}}
&1   &120& 74.5 & 82& 27& 37.5 \\
&5 & 600&98.5& 98.5& 65.5&  72\\
&10 &1200& 99.5& 99.5& 78& 84 \\
&30  & 3600&100& 100& 95.5&97.5\\
\hline
\end{tabular}
\end{table}

There are two common ways to check normality/Gaussianity, namely, graphical methods and numerical methods~\cite{razali2011power}. Here, the measurement $z(t,\omega)$, $t \in T$, $\omega \in \Omega$ is viewed as a stochastic process defined on the probability space $(\Omega, \mathcal F, P)$ over time $T$. The graphical method is implemented using the scaled error $\Delta e_s(\omega)$ along with the standard Gaussian distribution to visualize the deviation from Gaussianity. The numerical method is performed with the Shapiro-Wilk (SW) and Kolmogorov-Smirnov (KS) tests, which reject or accept null hypothesis of Gaussianity using test index and \% confidence index $\alpha$~\cite{razali2011power}. 
\begin{eqnarray}
\Delta e_s(\omega)=\frac{\Delta e (\omega)- \mu_{\Delta e}} {\sigma_{\Delta e}}, 
\end{eqnarray}
where $\mu_{\Delta e} $ and $\sigma_{\Delta e}$ are the mean and standard deviation of $\Delta e(\omega)$.

\begin{figure}[h]
\centering
\subfloat{\includegraphics[width=0.32\linewidth]{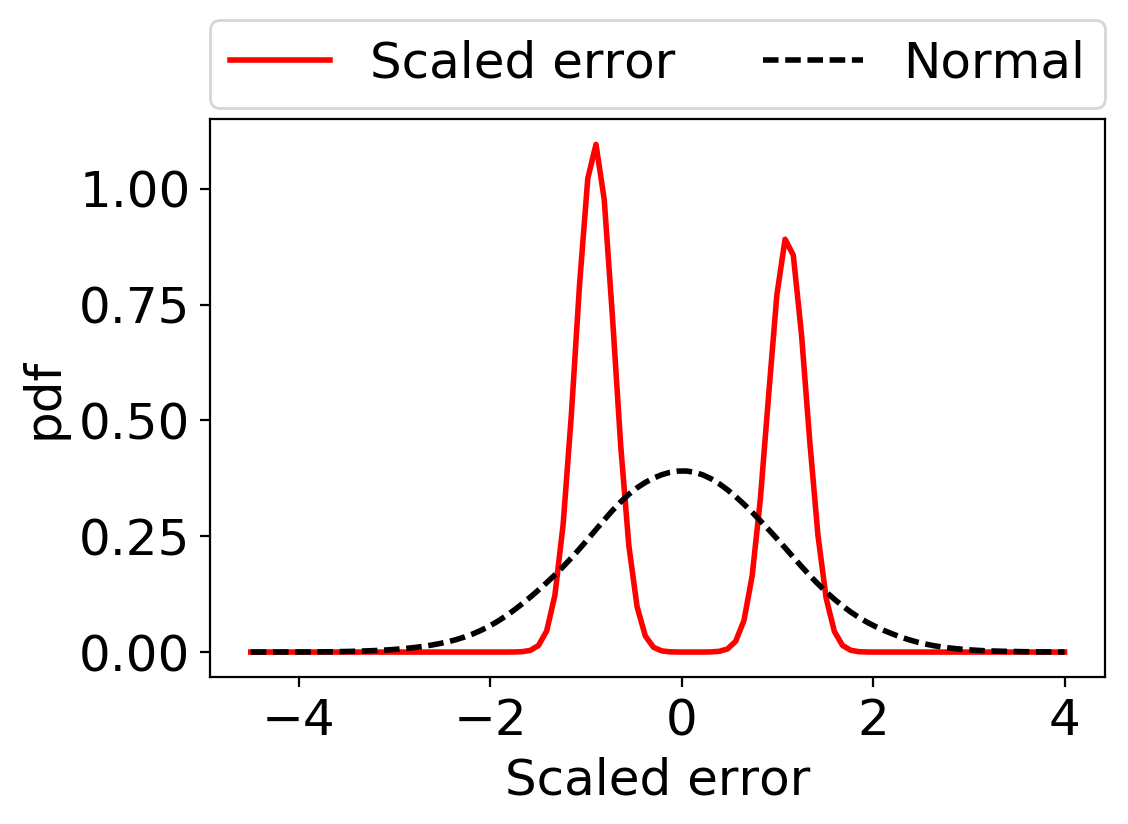}}\addtocounter{subfigure}{-1}
\subfloat[]{\includegraphics[width=0.32\linewidth]{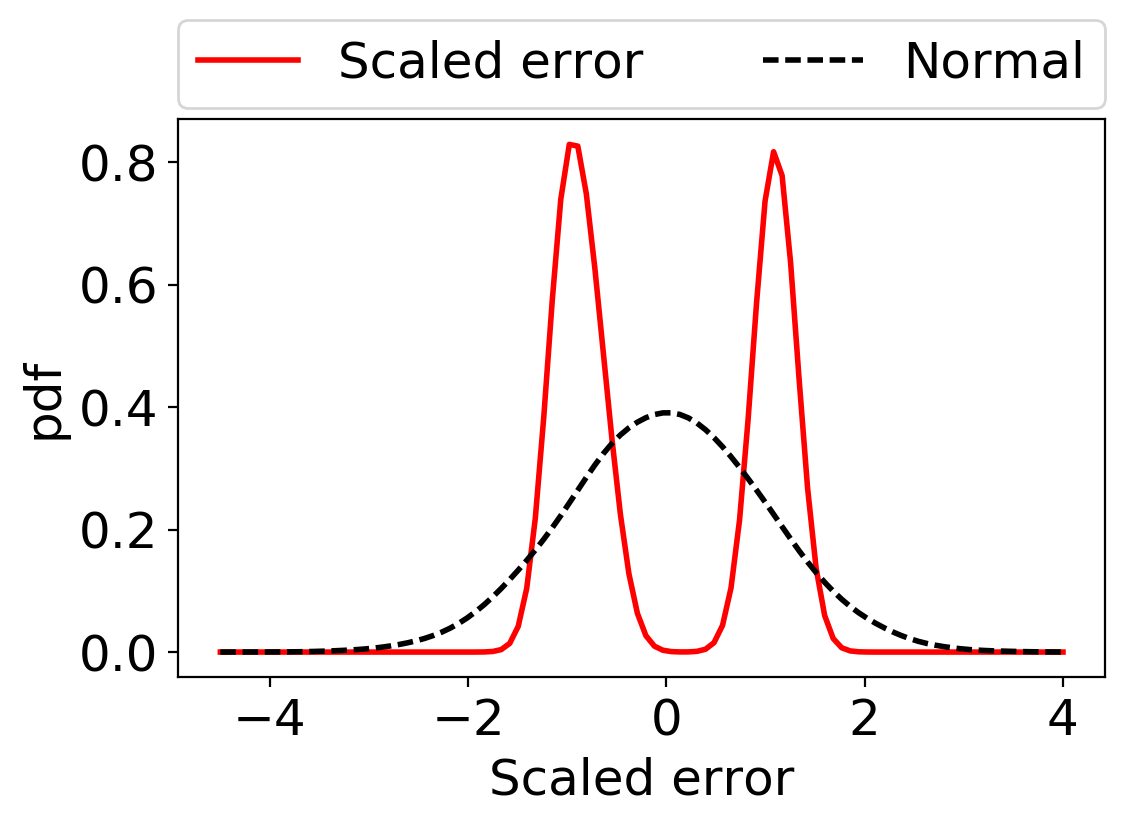}} 
\subfloat{\includegraphics[width=0.32\linewidth]{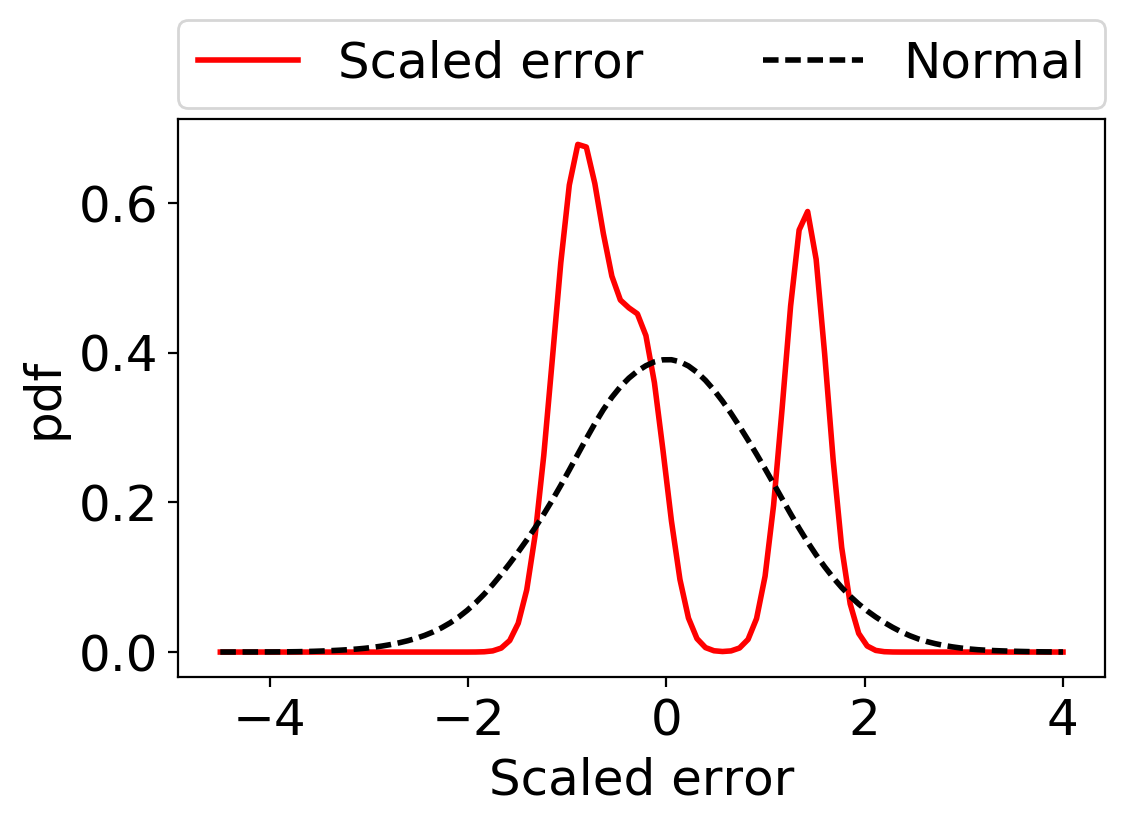}} \addtocounter{subfigure}{-1}\\
\vspace{-0.15in}
\subfloat{\includegraphics[width=0.32\linewidth]{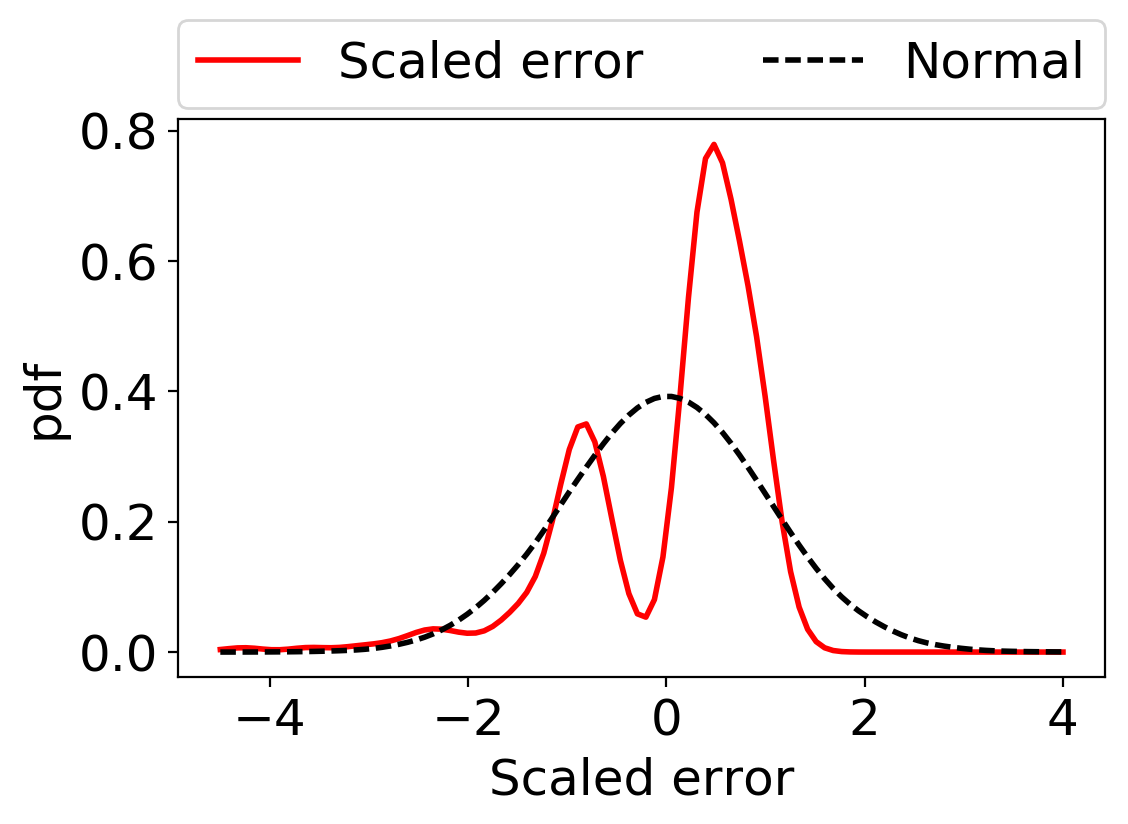}} \addtocounter{subfigure}{-1}
\subfloat[]{\includegraphics[width=0.32\linewidth]{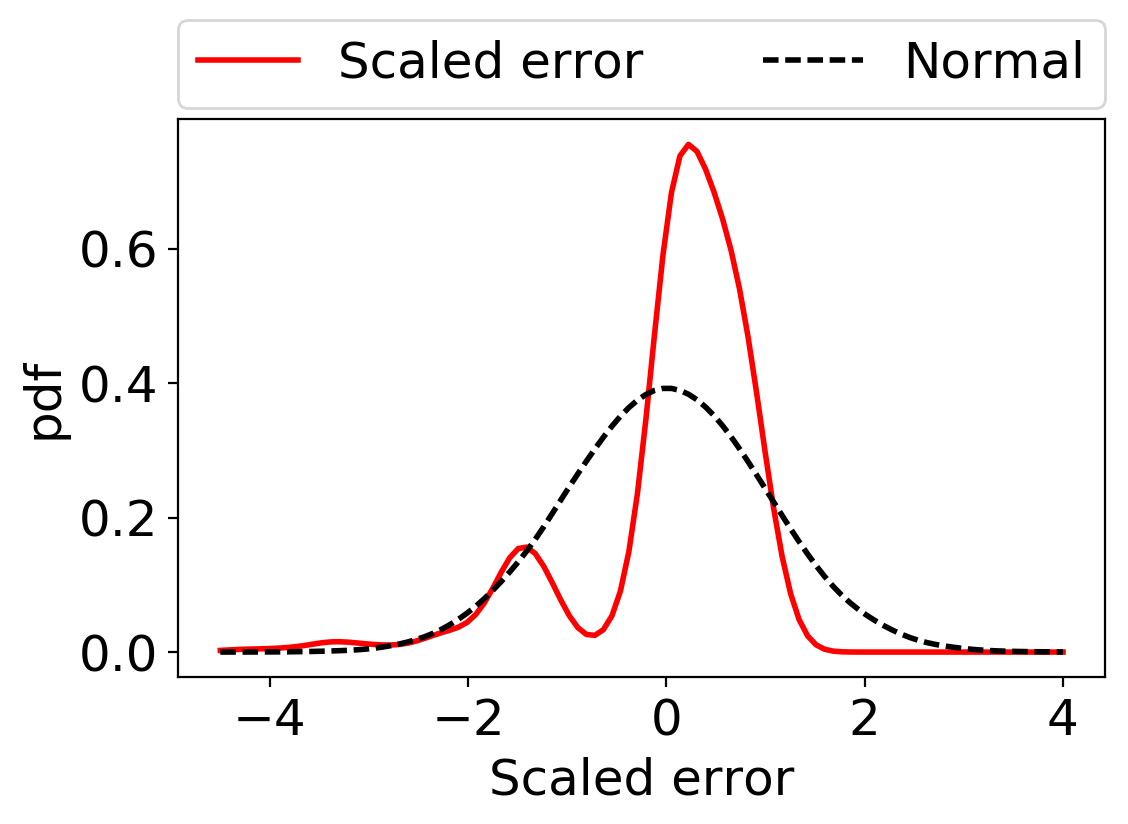}} \subfloat{\includegraphics[width=0.32\linewidth]{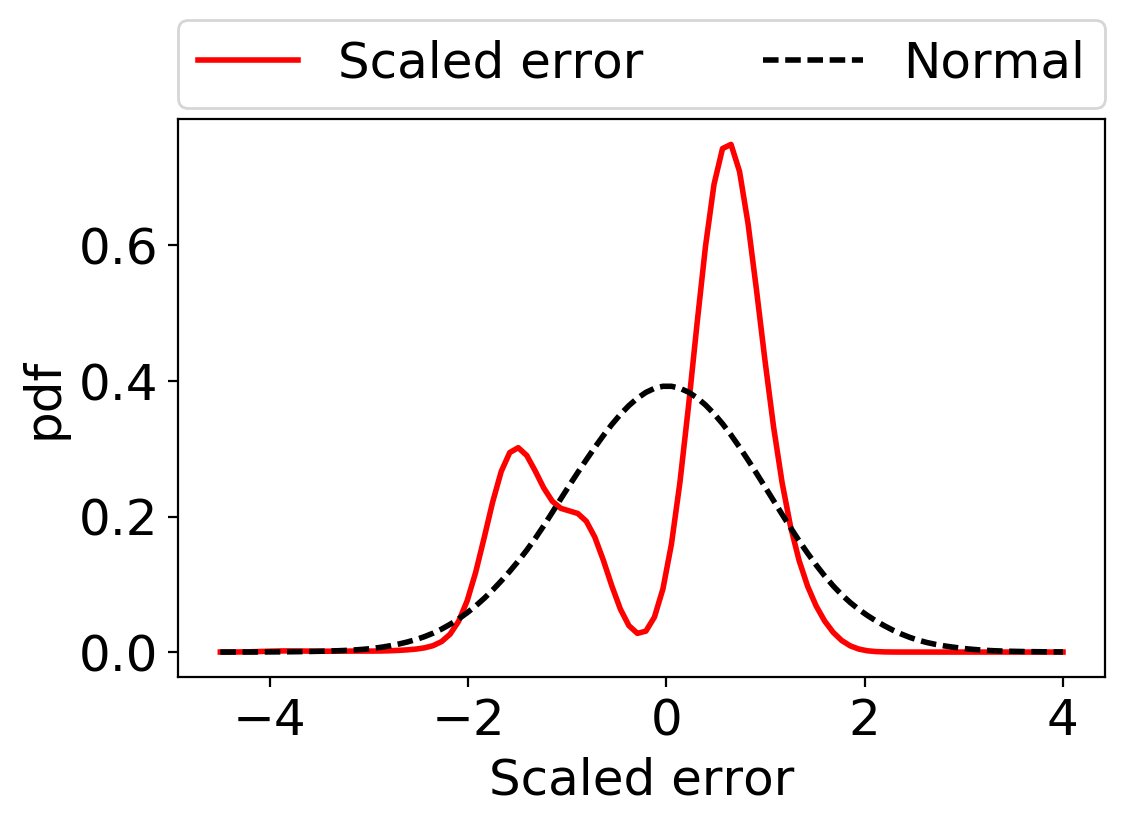}} \addtocounter{subfigure}{-1}\\
\vspace{-0.15in}
\subfloat[]{\includegraphics[width=0.5\linewidth]{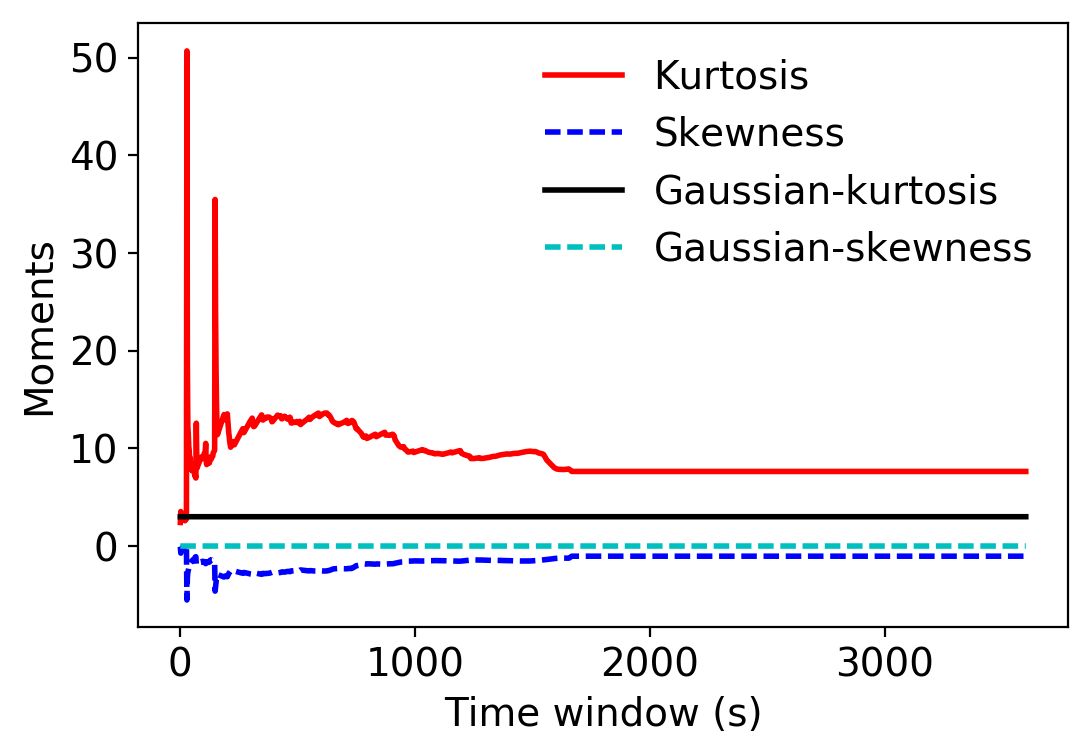}}
\subfloat[]{\includegraphics[width=0.5\linewidth]{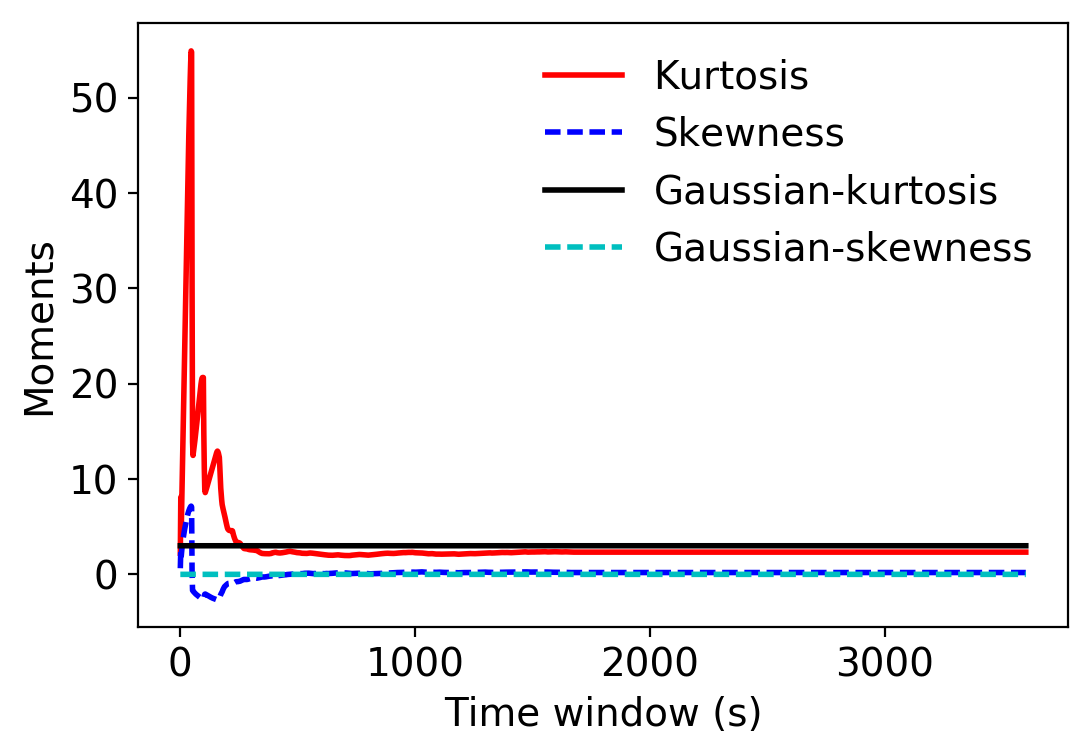}}
\caption{Test 2 results: (a) pdf of the voltage angle error differences $\Delta e_s$ with standard Gaussian for time windows of 1s, 5s, 10s; (b) pdf of the voltage magnitude error differences $\Delta e_s$ with standard Gaussian for time windows of 1s, 5s, and 10s; (c) varying  time windows versus skewness and kurtosis of voltage angle  $\Delta e$; and (d) varying update period versus skewness and kurtosis of voltage magnitude $\Delta e$.}\label{fig:pdf_philip}
\end{figure}
\begin{table}[h]
\centering
\caption{Summary of the Gaussianity Tests on PMU Data}\label{tab:Gper_philip}
\begin{tabular}{|c|c|c|c|c|c|c|}
\hline
\multicolumn{1}{|c|}{\multirow{3}{*}{\begin{tabular}[c]{@{}c@{}}Measurement\end{tabular}}}&
\multicolumn{1}{c|}{\multirow{3}{*}{\begin{tabular}[c]{@{}c@{}}Time\\ window\\ (s)\end{tabular}}} &\multicolumn{1}{c|}{\multirow{3}{*}{\begin{tabular}[c]{@{}c@{}}Sample\\size \end{tabular}}} & \multicolumn{4}{c|}{\begin{tabular}[c]{@{}c@{}}\% of non-Gaussian distributions\end{tabular}}                                   \\ \cline{4-7} 
& & & \multicolumn{2}{c|}{Shapiro-Wilk}& \multicolumn{2}{c|}{Kolmogorov-Smirnov}\\ \cline{4-7} 
& & & \multicolumn{1}{c|}{$\alpha$=5\%} & \multicolumn{1}{c|}{$\alpha$=10\%} & \multicolumn{1}{c|}{$\alpha$=5\%} & \multicolumn{1}{c|}{$\alpha$=10\%} \\ 
\hline
\multicolumn{1}{|c|}{\multirow{4}{*}{\begin{tabular}[c]{@{}c@{}}Voltage\\angle\end{tabular}}}
&1 &60&  44& 48& 24 & 28\\
&5& 300&96& 100& 88& 88 \\
&10 & 600&100 & 100& 100&100\\
&60 &3600&100 & 100& 100& 100\\
\hline
\multicolumn{1}{|c|}{\multirow{4}{*}{\begin{tabular}[c]{@{}c@{}}Voltage\\magnitude\end{tabular}}}
&1 & 60& 4& 8& 0 & 0\\
&5&300& 84& 88& 36& 40 \\
&10 &600& 92 & 92& 72&72\\
&60 &3600&100 & 100& 100& 100\\
\hline
\end{tabular}
\end{table}

\section{Results}

\begin{figure}[h]
\centering
\subfloat{\includegraphics[width=0.32\linewidth]{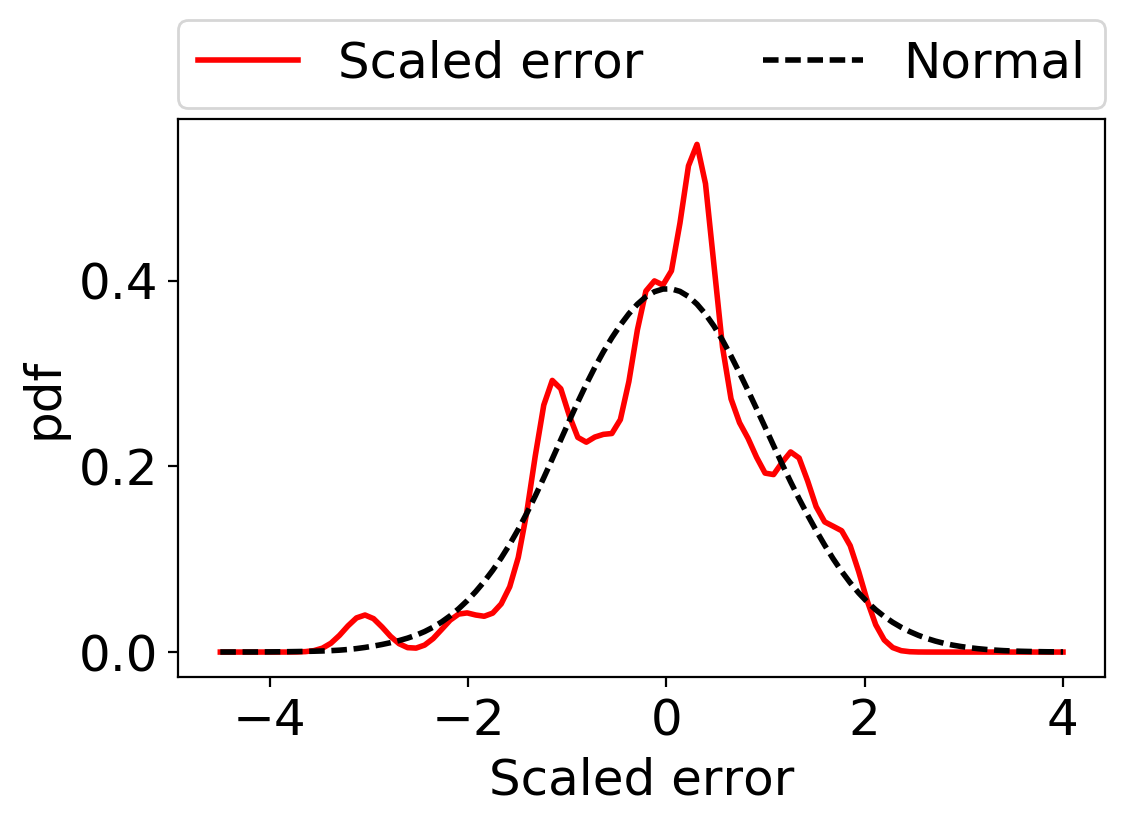}}\addtocounter{subfigure}{-1}
\subfloat[]{\includegraphics[width=0.32\linewidth]{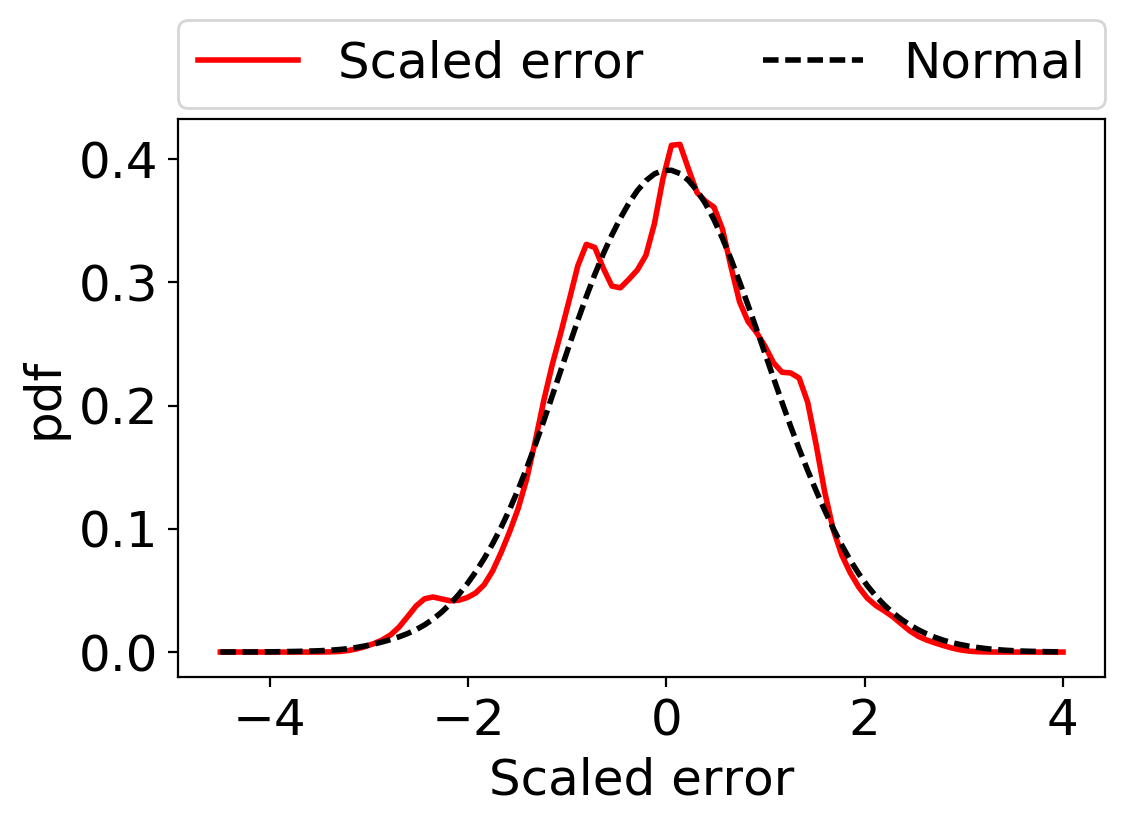}} 
\subfloat{\includegraphics[width=0.32\linewidth]{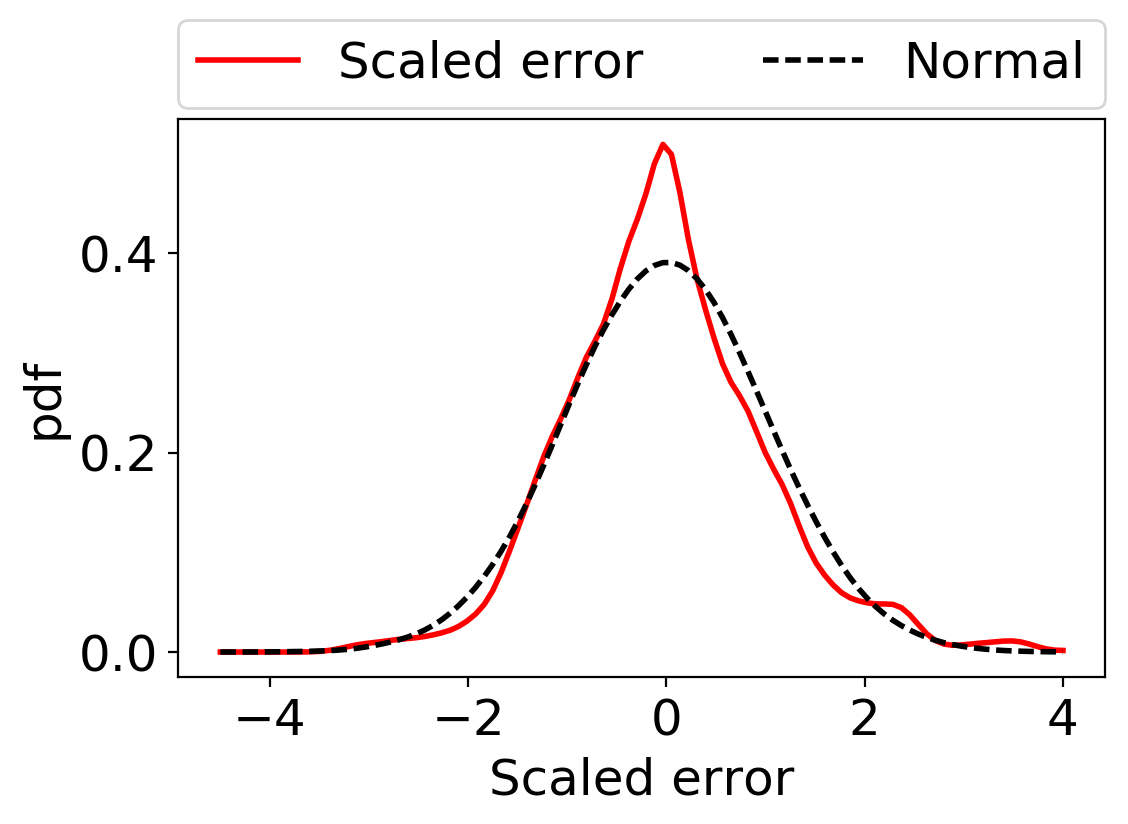}}\addtocounter{subfigure}{-1}\\
\vspace{-0.15in}
\subfloat{\includegraphics[width=0.32\linewidth]{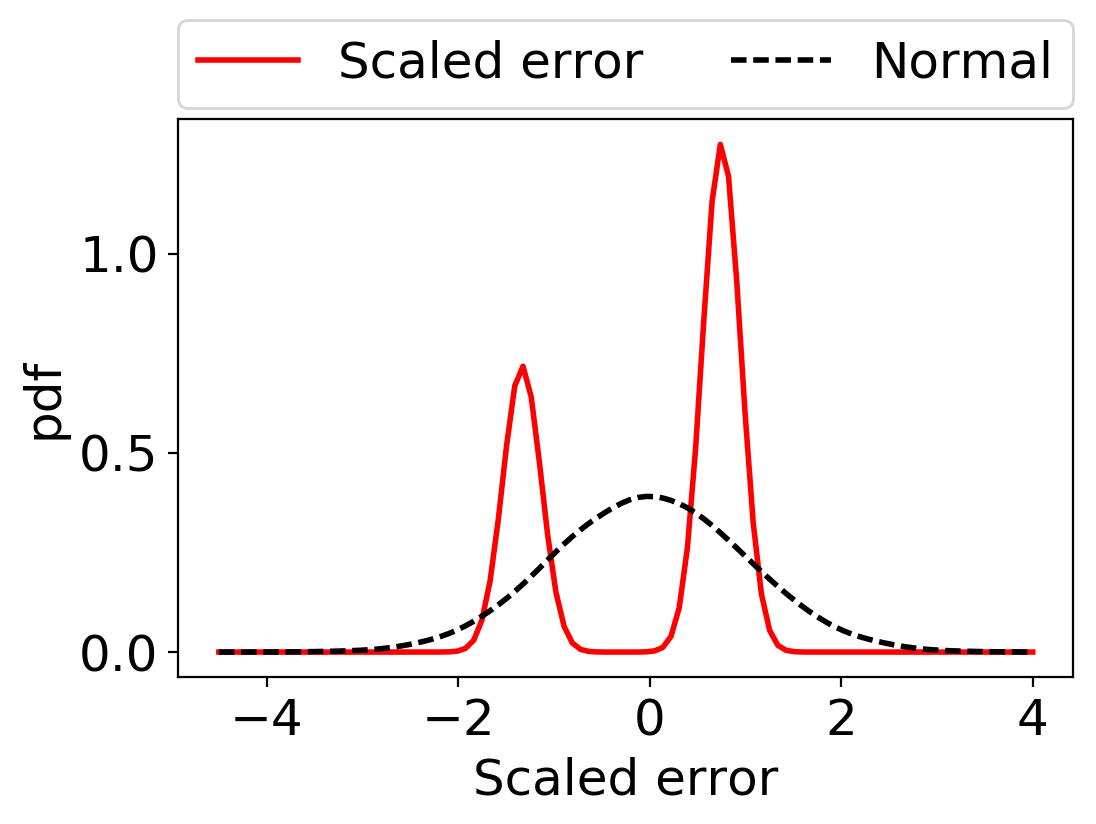}}\addtocounter{subfigure}{-1}
\subfloat[]{\includegraphics[width=0.32\linewidth]{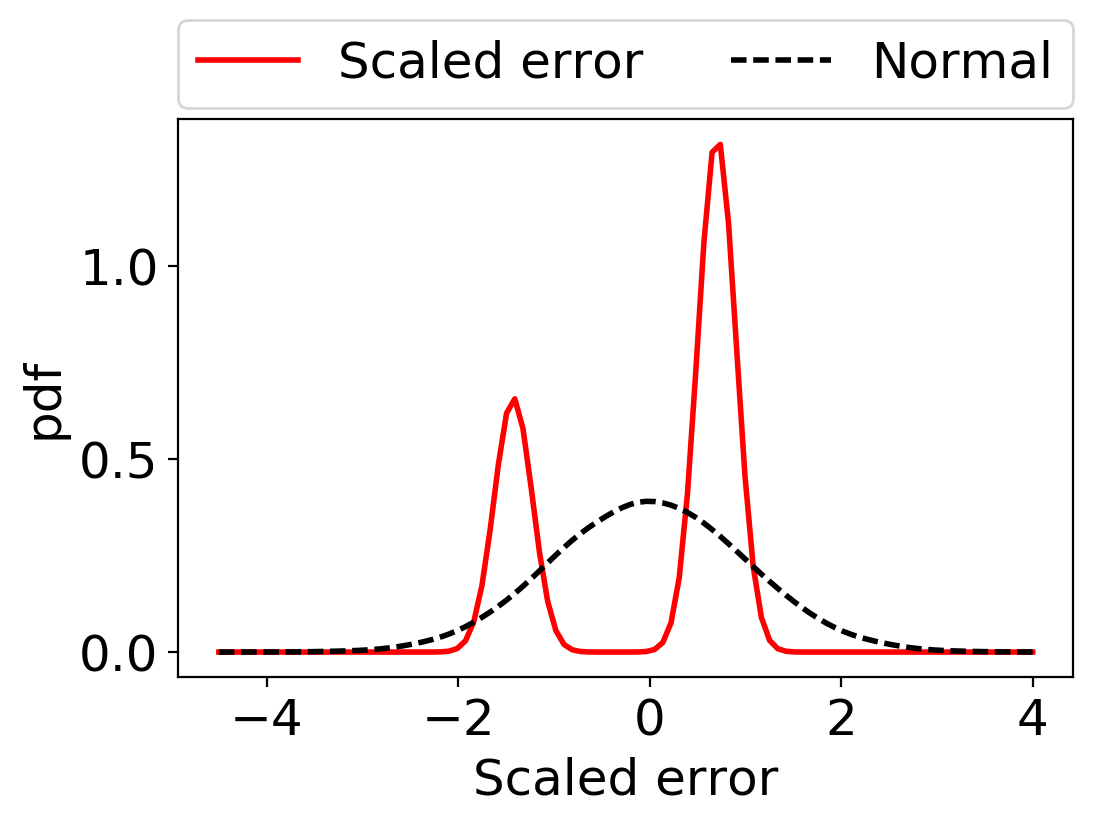}} 
\subfloat{\includegraphics[width=0.32\linewidth]{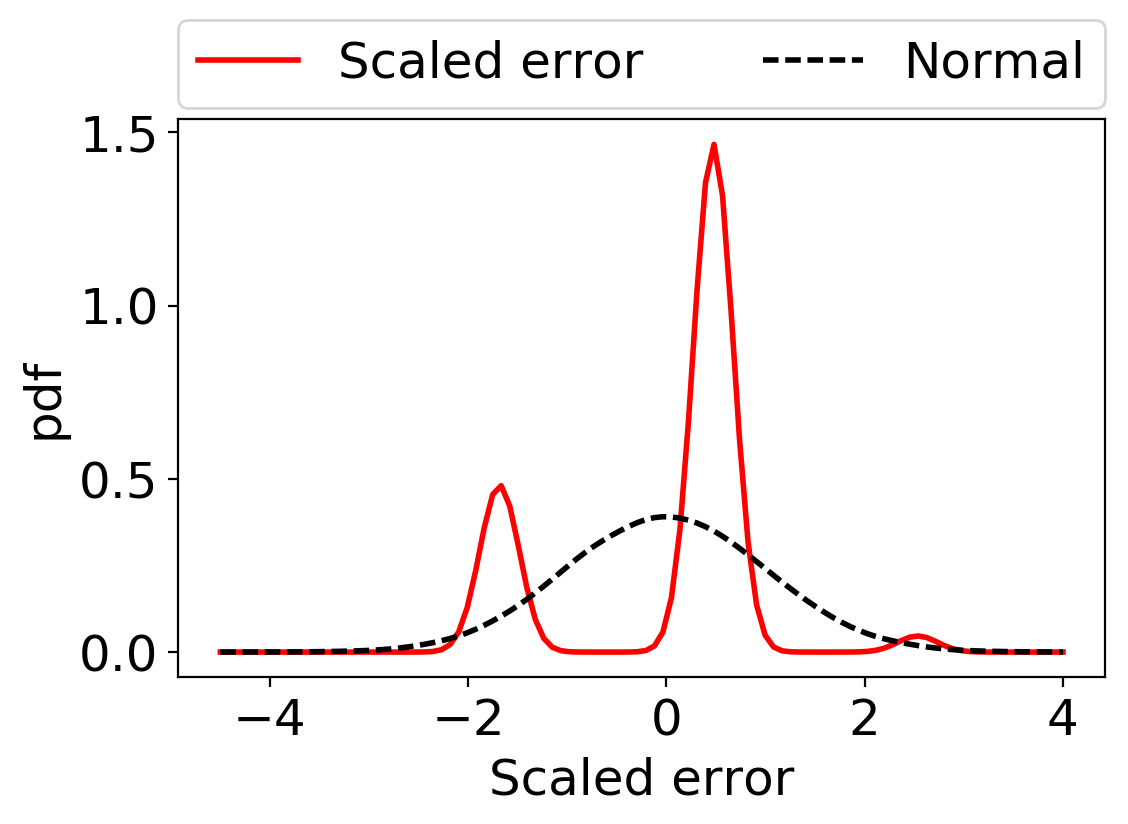}}\addtocounter{subfigure}{-1}\\
\vspace{-0.15in}
\subfloat[]{\includegraphics[width=0.5\linewidth]{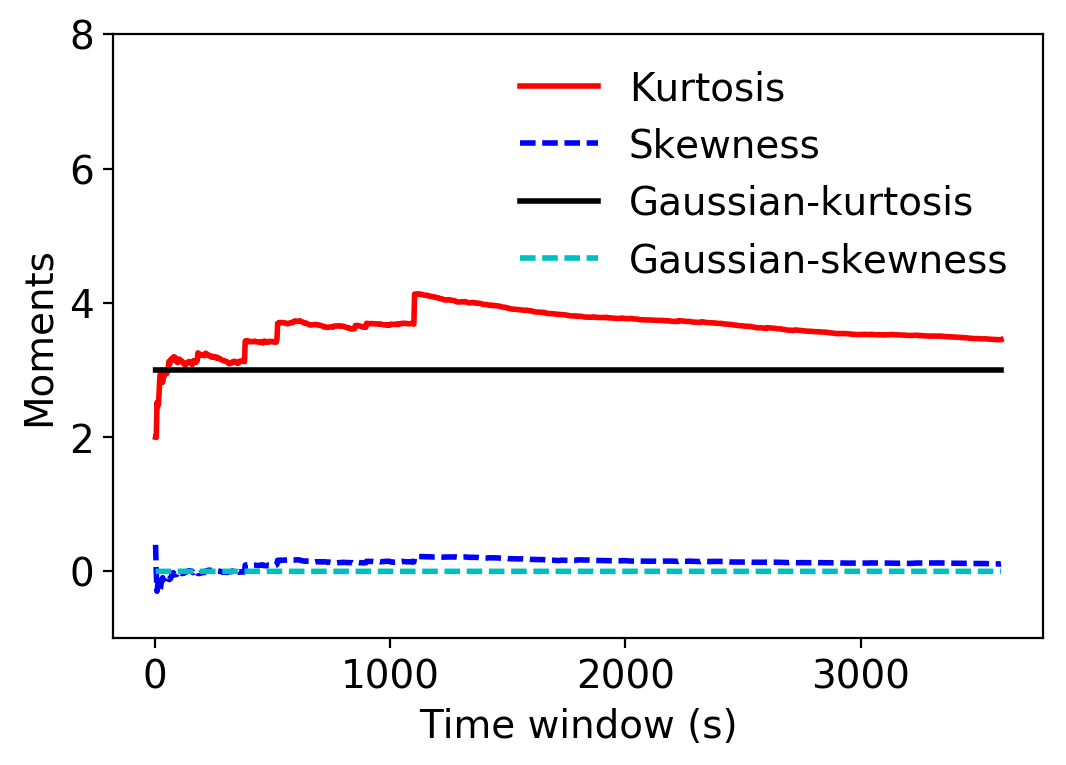}}
\subfloat[]{\includegraphics[width=0.5\linewidth]{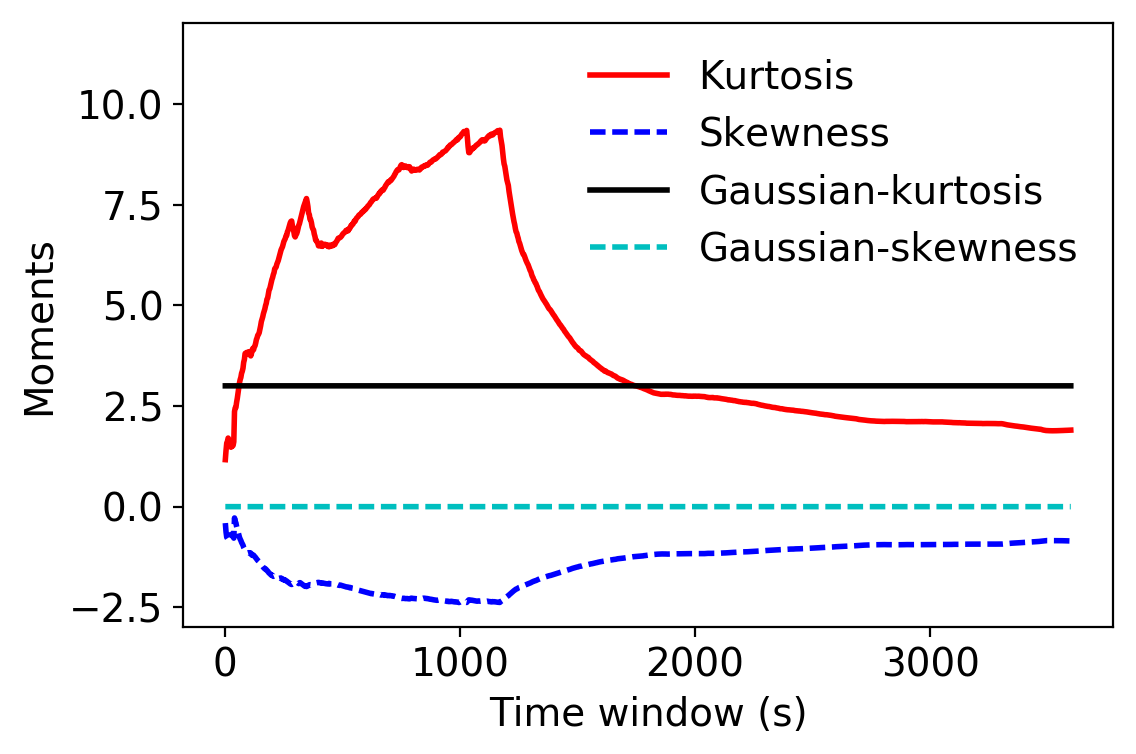}}
\caption{Test 3 results: (a) pdf of the voltage angle errors with standard Gaussian for time windows of 5s, 30s, 60s; (b) pdf of the voltage magnitude errors with standard Gaussian for time windows of 5s, 30s, and 60s; (c) varying time windows vs skewness and kurtosis of voltage angle $\Delta e$; and (d) varying time windows vs skewness and kurtosis of voltage magnitude $\Delta e$.}\label{fig:pdf_fru}
\end{figure}

\begin{table}[h]
\centering
\caption{Summary of the Gaussianity Tests on FDR Data}\label{tab:Gper}
\begin{tabular}{|c|c|c|c|c|c|c|}
\hline
\multicolumn{1}{|c|}{\multirow{3}{*}{\begin{tabular}[c]{@{}c@{}}Measurement\end{tabular}}}&
\multicolumn{1}{c|}{\multirow{3}{*}{\begin{tabular}[c]{@{}c@{}}Time\\ window\\ (s)\end{tabular}}} &\multicolumn{1}{c|}{\multirow{3}{*}{\begin{tabular}[c]{@{}c@{}}Sample\\size \end{tabular}}} & \multicolumn{4}{c|}{\begin{tabular}[c]{@{}c@{}}\% of non-Gaussian distributions\end{tabular}}                                   \\ \cline{4-7} 
& & & \multicolumn{2}{c|}{Shapiro-Wilk}& \multicolumn{2}{c|}{Kolmogorov-Smirnov}\\ \cline{4-7} 
& & & \multicolumn{1}{c|}{$\alpha$=5\%} & \multicolumn{1}{c|}{$\alpha$=10\%} & \multicolumn{1}{c|}{$\alpha$=5\%} & \multicolumn{1}{c|}{$\alpha$=10\%} \\ 
\hline
\multicolumn{1}{|c|}{\multirow{4}{*}{\begin{tabular}[c]{@{}c@{}}Voltage\\angle\end{tabular}}}
&5 &50& 32& 46& 4 & 10\\
&30&300& 100& 100& 100& 100 \\
&60 &600& 100& 100& 100& 100\\
&120 &1200& 100 & 100& 100& 100\\
\hline
\multicolumn{1}{|c|}{\multirow{4}{*}{\begin{tabular}[c]{@{}c@{}}Voltage\\magnitude\end{tabular}}}
&5 &50& 100& 100& 100 & 100\\
&30&300& 100& 100& 100& 100 \\
&60 &600& 100& 100& 100& 100\\
&120 &1200&100 & 100& 100& 100\\
\hline
\end{tabular}
\end{table}

In power engineering, distribution systems are monitored by various sensing and measuring devices with multiple time scales, like $\mu$PMU with reporting rate 120 fps, PMU with reporting rate 10/30/60 fps, and supervisory control and data acquisition (SCADA) updating a frame in every few seconds. In statistics, a statistical test has little power when the sample size is extremely small or large (e.g., sample size $<30$ or $>6000$)~\cite{razali2011power, ghasemi2012normality}. Thus, in the tests described below, the distributions under different time windows and sample sizes are considered collectively.

In Test 1, a voltage phasor in a primary distribution system is measured using two $\mu$PMUs, and in Test 2, a voltage phasor in a secondary distribution system is metered with two PMUs manufactured by the same vendor. Figs. \ref{fig:pdf}(a), \ref{fig:pdf}(b), \ref{fig:pdf_philip}(a), and \ref{fig:pdf_philip}(b) describe the probability density functions (pdf) of the scaled error $\Delta e_s$ along with the standard Gaussian distribution within different time windows. The results show that the distributions in various cases are all non-Gaussian with multiple peaks and they are non-symmetric. Figs. 1(c) 1(d), 2(c), and 2(d) show the varying time window versus skewness (third-order moment) and kurtosis (fourth-order moment) of $\Delta e$. We observe that the skewness deviates from zero and kurtosis departs from Gaussian-kurtosis, which demonstrate the non-symmetric and long- or short-tail nature of the pdf of $\Delta e$ even for the large time window. Also, Tables I and II give the percentage of non-Gaussian random variables (rvs) out of the total random samples taken with different time windows and different sample sizes (note that SW test is generally more powerful than KS test and the power of both KS and SW tests is low for small sample, e.g, sample size $<60$ ~\cite{razali2011power}). It is observed that most rvs are non-Gaussian based on the SW and KS tests, and that their joint distribution is non-Gaussian. Moreover, the skewness and kurtosis in Figs. 2(c) and 2(d) are much higher than the ones in Figs. 1(c) and 1(d), partially because the measurement error in the secondary distribution system is more vulnerable to environmental changes than the measurement error in the primary distribution system. 

Furthermore, in Test 3, an off-line test is carried out on the FNET/GridEye platform using the FDR with the reporting rate of 10 fps. It is found from Fig. \ref{fig:pdf_fru} and Table III that even though there are few environmental fluctuations in the laboratory environment, the observed measurement errors still follow a non-Gaussian distribution. Through the on-line and off-line measurements and the graphical and numerical analysis, extensive results reveal that the real-world measurement error potentially follows a non-Gaussian distribution.

\section{Conclusion}

Today's electric power distribution system is being transformed from a passive system into an active and intelligent network. It is advantageous to understand the distribution-level measurement characteristics, which is critical to distribution system planning and operation. This letter studies the distribution-level synchrophasor measurement error and shows that, based on a series of tests, the measurement error follows a non-Gaussian distribution instead of the traditionally-assumed Gaussian distribution. It suggests the use of non-Gaussian or Gaussian mixture model (GMM) for modeling the distribution synchrophasor measurement error, which is more accurate and more realistic than the traditional Gaussian model. The presented measurements and analysis will become helpful for the understanding of distribution measurement  characteristics, and for the modeling and simulation of distribution system applications. 

The future work includes parameterization of both synchronized and non-synchronized distribution measurements with advanced data analytics and data-intensive machine learning, and also the application of the results to DER and DMS studies.

\section{Acknowledgment}
This work was sponsored by the U.S. Department of Energy through its Laboratory Directed Research and Development (LDRD) program. This work was performed under the auspices of the U.S. Department of Energy by Lawrence Livermore National Laboratory under Contract DE-AC52-07NA27344 with IM release number LLNL-JRNL-740270.
\bibliographystyle{IEEEtran}
\bibliography{ref}

\end{document}